\documentclass[10pt,a4paper]{article}
\usepackage[utf8]{inputenc}

\usepackage{grffile}  %% deals with extension problems.
\usepackage{hyperref}  %% links
\usepackage{a4wide}

\usepackage{amsmath}
\usepackage{amsfonts}
\usepackage{amssymb}
\usepackage{mathtools}
\usepackage{natbib}

\title{Efficient Spherical Harmonic Transforms aimed at pseudo-spectral numerical simulations}

\author{Nathana\"el Schaeffer}
\author{Nathana\"el Schaeffer \\	{\small ISTerre, Universit\'e de Grenoble 1, CNRS, F-38041 Grenoble, France}}

\begin{document}

\maketitle

\begin{abstract}
In this paper, we report on very efficient algorithms for the spherical harmonic transform (SHT).
Explicitly vectorized variations of the algorithm based on the Gauss-Legendre quadrature are discussed and implemented in the \texttt{SHTns} library which includes scalar and vector transforms.
The main breakthrough is to achieve very efficient on-the-fly computations of the Legendre associated functions, even for very high resolutions, by taking advantage of the specific properties of the SHT and the advanced capabilities of current and future computers.
This allows us to simultaneously and significantly reduce memory usage and computation time of the SHT.
We measure the performance and accuracy of our algorithms.
Even though the complexity of the algorithms implemented in \texttt{SHTns} are in ${\cal O}(N^3)$ (where $N$ is the maximum harmonic degree of the transform), they perform much better than any third party implementation, including lower complexity algorithms, even for  truncations as high as $N=1023$.
\texttt{SHTns} is available at \url{https://bitbucket.org/nschaeff/shtns} as open source software.
\end{abstract}

\section{Introduction}

Spherical harmonics are the eigenfunctions of the Laplace operator on the 2-sphere.
They form a basis and are useful and convenient to describe data on a sphere in a consistent way in spectral space.
Spherical Harmonic Transforms (SHT) are the spherical counterpart of the Fourier transform, casting spatial data to the spectral domain and vice versa.
They are commonly used in various pseudo-spectral direct numerical simulations in spherical geometry, for simulating the Sun or the liquid core of the Earth among others \citep{glatzmaier84, sakuraba1999, dynamo_benchmark, brun2009, wicht10}.

All numerical simulations that take advantage of spherical harmonics use the classical Gauss-Legendre algorithm (see section \ref{sec:sht}) with complexity ${\cal O}(N^3)$ for a truncation at spherical harmonic degree $N$.
As a consequence of this high computational cost when $N$ increases, high resolution spherical codes currently spend most of their time performing SHT.
A few years ago, state of the art numerical simulations used $N=255$ \citep{sakuraba09}.

However, there exist several asymptotically fast algorithms \citep{driscoll94, Potts98, Mohlenkamp99, Suda2002Fast, healy03, tygert2008}, but the overhead for these fast algorithms is such that they do not claim to be effectively faster for $N < 512$. In addition, some of them lack stability (the error becomes too large even for moderate $N$) and flexibility (e.g. $N+1$ must be a power of 2).

Among the asymptotically fast algorithms, only two have open-source implementations, and the only one which seems to perform reasonably well is \texttt{SpharmonicKit}, based on the algorithms described by \citet{healy03}.
Its main drawback is the need of a latitudinal grid of size $2(N+1)$ while the Gauss-Legendre quadrature allows the use of only $N+1$ collocation points.
Thus, even if it were as fast as the Gauss-Legendre approach for the same truncation $N$, the overall numerical simulation would be slower because it would operate on twice as many points.
These facts explain why the Gauss-Legendre algorithm is still the most efficient solution for numerical simulations.

A recent paper \citep{dickson11} reports that carefully tuned software could finally run 9 times faster on the same CPU than the initial non-optimized version, and insists on the importance of vectorization and careful optimization of the code.
As the goal of this work is to speed-up numerical simulations, we have written a highly optimized and explicitly vectorized version of the Gauss-Legendre SHT algorithm.
The next section recalls the basics of spherical harmonic transforms.
We then describe the optimizations we use and we compare the performance of our transform to other SHT implementations.
We conclude this paper by a short summary and perspectives for future developments.

\section{Spherical Harmonic Transform (SHT)}
\label{sec:sht}

\subsection{Definitions and properties}

The orthonormalized spherical harmonics of degree $n$ and order $-n \leq m \leq n$ are functions defined on the sphere as:
\begin{equation}  \label{eq:ylm}
 Y_n^m(\theta, \phi) = P_n^m(\cos \theta) \, \exp(im\phi)
\end{equation}
where $\theta$ is the colatitude, $\phi$ is the longitude and $P_n^m$ are the associated Legendre polynomials normalized for spherical harmonics
\begin{equation}
 P_n^m (x) = (-1)^m \ \sqrt{\frac{2n+1}{4\pi}} \sqrt{\frac{(n-|m|)!}{(n+|m|)!}} \ (1-x^2)^{|m|/2}\ \frac{d^{|m|}}{dx^{|m|}}P_n(x)
\end{equation}
which involve derivatives of Legendre Polynomials $P_n(x)$ defined by the following recurrence:
\begin{eqnarray*}
	P_0(x) &=& 1	\\
	P_1(x) &=& x	\\
	n P_n(x) &=& (2n-1)\, x \, P_{n-1}(x) \: - \: (n-1) P_{n-2}(x)
\end{eqnarray*}

The spherical harmonics $Y_n^m(\theta,\phi)$ form an orthonormal basis for functions defined on the sphere:
\begin{equation}
\int_0^{2\pi} \int_0^{\pi} Y_n^m(\theta,\phi) \overline{Y_l^k}(\theta,\phi) \sin\theta \,d\theta \,d\phi = \delta_{nl} \delta_{mk}
\end{equation}
with $\delta_{ij}$ the Kronecker symbol, and $\overline{z}$ the complex conjugate of $z$.
By construction, they are eigenfunctions of the Laplace operator on the unit sphere:
\begin{equation}
	\Delta Y_n^m = -n(n+1) Y_n^m
\end{equation}
This property is very appealing for solving many physical problems in spherical geometry involving the Laplace operator.

\subsection{Synthesis or inverse transform}

The Spherical Harmonic synthesis is the evaluation of the sum
\begin{equation}
 f(\theta,\phi) = \sum_{n=0}^N \sum_{m=-n}^{n} f_n^m Y_n^m(\theta, \phi) 
\end{equation}
up to degree $n=N$, given the complex coefficients $f_n^m$.
If $ f(\theta,\phi)$ is a real-valued function, $f_n^{-m} = \overline{f_n^m}$.

The sums can be exchanged, and using the expression of $Y_n^m$ we can write
\begin{equation}
 f(\theta,\phi) = \sum_{m=-N}^{N} \left( \sum_{n=|m|}^N  f_n^m P_n^m(\cos\theta) \right) e^{im\phi}
\end{equation}
From this last expression, it appears that the summation over $m$ is a regular Fourier Transform.
Hence the remaining task is to evaluate
\begin{equation} \label{eq:synth_direct}
 f_m(\theta) = \sum_{n=|m|}^N f_n^m P_n^m(\cos\theta)
\end{equation}
or its discrete version at given collocation points $\theta_j$.

\subsection{Analysis or forward transform}

The analysis step of the SHT consists in computing the coefficients
\begin{equation}
 f_n^m = \int_0^{2\pi} \int_0^{\pi} f(\theta,\phi) \overline{Y_n^m}(\theta,\phi) \sin\theta \,d\theta \,d\phi
\end{equation}
The integral over $\phi$ is obtained using the Fourier Transform:
\begin{equation}
 f_m(\theta) = \int_0^{2\pi} f(\theta,\phi) e^{-im\phi}\, d\phi
\end{equation}
so the remaining Legendre transform reads
\begin{equation}	\label{eq:analysis}
 f_n^m = \int_0^{\pi} f_m(\theta) P_n^m(\cos\theta) \sin\theta \,d\theta
\end{equation}
The discrete problem reduces to the appropriate quadrature rule to evaluate the integral (\ref{eq:analysis}) knowing only the values $f_m(\theta_j)$.
In particular, the use of the Gauss-Legendre quadrature replaces the integral of expression \ref{eq:analysis} by the sum
\begin{equation}	\label{eq:gauss}
 f_n^m = \sum_{j=1}^{N_\theta} f_m(\theta_j) P_n^m(\cos\theta_j) w_j
\end{equation}
where $\theta_j$ and $w_j$ are respectively the Gauss nodes and weights \citep{dlmf_gauss}. Note that the sum equals the integral if $f_m(\theta) P_n^m(\cos\theta)$ is a polynomial in $\cos\theta$ of order $2N_\theta-1$ or less.
If $f_m(\theta)$ is given by expression \ref{eq:synth_direct}, then $f_m(\theta) P_n^m(\cos\theta)$ is always a polynomial in $\cos\theta$, of degree at most $2N$. Hence the Gauss-Legendre quadrature is exact for $N_\theta \geq N+1$.

A discrete spherical harmonic transform using Gauss nodes as latitudinal grid points and a Gauss-Legendre quadrature for the analysis step is referred to as a Gauss-Legendre algorithm.

\section{Optimization of the Gauss-Legendre algorithm}

\subsection{Standard optimizations}

Let us first recall some standard optimizations found in almost every serious implementation of the Gauss-Legendre algorithm.
All the following optimizations are used in the \texttt{SHTns} library.

\paragraph{Use the Fast-Fourier Transform}
The expressions of section \ref{sec:sht} show that part of the SHT is in fact a Fourier transform. The fast Fourier transform (FFT) should be used for this part, as it improves accuracy and speed. \texttt{SHTns} uses the \texttt{FFTW} library\citep{fftw}, a portable, flexible and highly efficient FFT implementation.

\paragraph{Take advantage of Hermitian symmetry for real data} When dealing with real-valued data, the spectral coefficients fulfill $f_n^{-m} = \overline{f_n^{m}}$, so we only need to store them for $m \geq 0$. This also allows the use of faster real-valued FFTs.
% \texttt{SHTns} only supports real-valued spatial fields.

\paragraph{Take advantage of mirror symmetry}
Due to the defined symmetry of spherical harmonics with respect to a reflection about the equator
$$P_n^m(\cos(\pi - \theta)) = (-1)^{n+m} \, P_n^m(\cos\theta)$$
one can reduce by a factor of 2 the operation count of both forward and inverse transforms.

\begin{figure}
\centering
\includegraphics[width=0.6\textwidth]{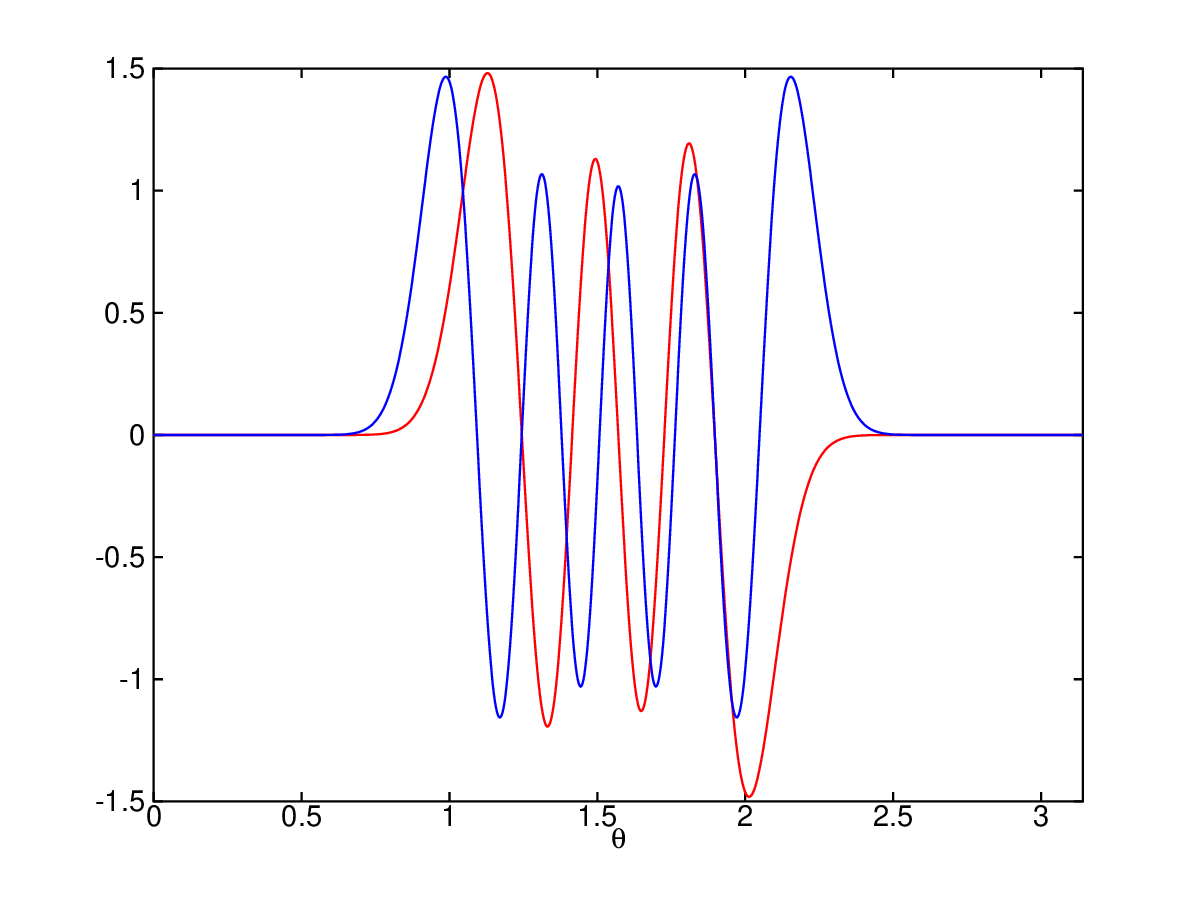}
\caption{Two associated Legendre polynomials of degree $n=40$ and order $m=33$ (blue) and $m=36$ (red), showing the localization near the equator.}
\label{fig:polar}
\end{figure}

\paragraph{Precompute values of $P_n^m$}
The coefficients $P_n^m(\cos\theta_j)$ appear in both synthesis and analysis expressions (\ref{eq:synth_direct} and \ref{eq:analysis}), and can be precomputed and stored for all ($n$,$m$,$j$).
When performing multiple transforms, it avoids computing the Legendre polynomial recursion at every transform and saves some computing power, at the expense of memory bandwidth.
This may or may not be efficient, as we will discuss later.

\paragraph{Polar optimization} High order spherical harmonics have their magnitude decrease exponentially when approaching the poles as shown in Figure \ref{fig:polar}.
Hence, the integral of expression \ref{eq:analysis} can be reduced to
\begin{equation}
 f_n^m = \int_{\theta_0^{mn}}^{\pi-\theta_0^{mn}} f_m(\theta) P_n^m(\cos\theta) \sin\theta \,d\theta
\end{equation}
where $\theta_0^{mn} \ge 0$ is a threshold below which $P_n^m$ is considered to be zero.
Similarly, the synthesis of $f_m(\theta)$ (eq. \ref{eq:synth_direct}) is only needed for $\theta_0^{mn} \le \theta \le \pi - \theta_0^{mn}$.
\texttt{SHTns} uses a threshold $\theta_0^{mn}$ that does not depend on $n$, which leads to around 5\% to 20\% speed increase, depending on the desired accuracy and the truncation $N$. %especially at high truncation $N$.

\subsection{On-the-fly algorithms and vectorization}
\label{sec:fly}

It can be shown that $P_n^m(x)$ can be computed recursively by
\begin{eqnarray}
	P_m^m(x) &=& a_m^m \, \left( 1-x^2 \right)^{|m|/2}   \\
	P_{m+1}^m(x)  &=& a_{m+1}^m \, x P_m^m(x)  \\ \label{eq:rec}
	P_n^m(x) &=& a_n^m \, x P_{n-1}^m(x) + b_n^m \, P_{n-2}^m(x) 
\end{eqnarray}
with
\begin{eqnarray}
	a_m^m &=&  \sqrt{ \frac{1}{4\pi} \: \prod_{k=1}^{|m|} \frac{2k+1}{2k} } \\	%	 a_m^m &= \frac{1}{\sqrt{4\pi}} \, \frac{\sqrt{(2m+1)!}}{m! \: 2^m} &
	a_n^m &=&  \sqrt{\frac{4n^2-1}{n^2-m^2}} \\
	b_n^m &=& -\sqrt{\frac{2n+1}{2n-3} \: \frac{(n-1)^2-m^2}{n^2-m^2}} 
\end{eqnarray}
The coefficients $a_n^m$ and $b_n^m$ do not depend on $x$, and can be easily precomputed and stored into an array of $(N+1)^2$ values.
This has to be compared to the order $N^3$ values of $P_n^m(x_j)$, which are usually precomputed and stored in the spherical harmonic transforms implemented in numerical simulations.
The amount of memory required to store all $P_n^m(x_j)$ in double-precision is at least $2(N+1)^3$ bytes, which gives 2Gb for $N=1023$.
Our on-the-fly algorithm only needs about $8(N+1)^2$ bytes of storage (same size as a spectral representation $f_n^m$), that is 8Mb for $N=1023$.
When $N$ becomes very large, it is no longer possible to store $P_n^m(x_j)$ in memory (for $N \gtrsim 1024$ nowadays) and on-the-fly algorithms (which recompute $P_n^m(x_j)$ from the recurrence relation when needed) are then the only possibility.

We would like to stress that even far from that storage limit, on-the-fly algorithm can be significantly faster thanks to vector capabilities of modern processors.
Most desktop and laptop computers, as well as many high performance computing clusters, have support for Single-Instruction-Multiple-Data (SIMD) operations in double precision. The SSE2 instruction set is available since year 2000 and currently supported by almost every PC, allowing to perform the same double precision arithmetic operations on a vector of 2 double precision numbers, effectively doubling the computing power.
The recently introduced AVX instruction set increases the vector size to 4 double precision numbers.
This means that $P_n^m(x)$ can be computed from the recursion relation \ref{eq:rec} (which requires 3 multiplications and 1 addition) for 2 or 4 values of $x$ simultaneously, which may be faster than loading pre-computed values from memory.
Hence, as already pointed out by \citet{dickson11}, it is therefore very important to use the vector capabilities of modern processors to address their full computing power.
Furthermore, when running multiple transforms on the different cores of a computer, the performance of on-the-fly transforms (which use less memory bandwidth) scales much better than algorithms with precomputed matrices, because the memory bandwidth is shared between cores.
Superscalar architectures that do not have double-precision SIMD instructions but have many computation units per core (like the POWER7 or SPARC64) could also benefit from on-the-fly transforms by saturating the many computation units with independent computations (at different $x$).

\begin{figure*}
\centering
\includegraphics[width=0.75\textwidth]{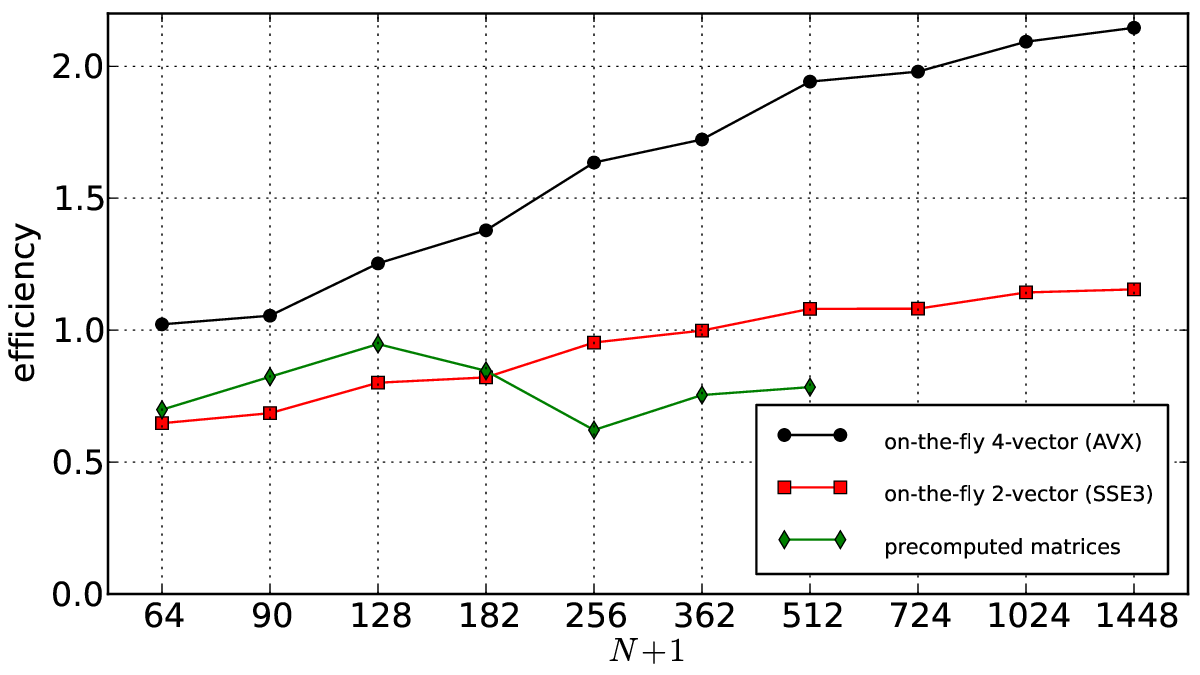}
\caption{Efficiency $(N+1)^3/(2tf)$ of various algorithms, where $t$ is the execution time and $f$ the frequency of the Xeon E5-2680 CPU (2.7GHz).
On-the-fly algorithms with two different vector sizes are compared with the algorithm using precomputed matrices.
Note the influence of hardware vector size for on-the-fly algorithms (AVX vectors pack 4 double precision floating point numbers where SSE3 vectors pack only 2).
The efficiency of the algorithm based on precomputed matrices drops above $N=127$ probably due to cache size limitations.}
\label{fig:avx}
\end{figure*}

Figure \ref{fig:avx} shows the benefit of explicit vectorization of on-the-fly algorithms on an intel Xeon E5-2680 (\emph{Sandy Bridge} architecture with AVX instruction set running at 2.7GHz) and compares on-the-fly algorithms with algorithms based on precomputed matrices.
With the 4-vectors of AVX, the fastest algorithm is always on-the-fly, while for 2-vectors, the fastest algorithm uses precomputed matrices for $N \lesssim 200$.
In the forthcoming years, wider vector architecture are expected to become widely available, and the benefits of on-the-fly vectorized transforms will become even more important.

\paragraph{Runtime tuning}

We have now two different available algorithms: one uses precomputed values for $P_n^m(x)$ and the other one computes them on-the-fly at each transform.
The \texttt{SHTns} library compares the time taken by those algorithms (and variants) at startup and chooses the fastest, similarly to what the \texttt{FFTW} library\citep{fftw} does.
The time overhead required by runtime tuning can be several order of magnitude larger than that of a single transform.
The observed performance gain varies between 10 and 30\%.
This is significant for numerical simulations, but runtime tuning can be entirely skipped for applications performing only a few transforms, in which case there is no noticeable overhead.

\subsection{Multi-threaded transform}

Modern computers have several computing cores.
We use OpenMP to implement a multi-threaded algorithm for the Legendre transform including the above optimizations and the \emph{on-the-fly} approach.
The lower memory bandwidth requirements for the \emph{on-the-fly} approach is an asset for a multi-threaded transform because if each thread would read a different portion of a large matrix, it can saturate the memory bus very quickly.
The multi-threaded Fourier transform is left to the FFTW library.

We need to decide how to share the work between different threads.
Because we compute the $P_n^m$ on the fly using the recurrence relation \ref{eq:rec}, we are left with 
each thread computing different $\theta$, or different $m$.
As the analysis step involve a sum over $\theta$, we choose the latter option.

From equation \ref{eq:synth_direct}, we see that the number of terms involved in the sum depends on $m$, so that the computing cost will also depend on $m$. In order to achieve the best workload balance between a team of $p$ threads, the thread number $i$ ($0 \leq i < p$) handles $m=i+kp \leq N$, with integer $k$ from $0$ to $(N+1)/p$.

For different thread number $p$, we have measured the time $T_s(p)$ and $T_a(p)$ needed for a scalar spherical harmonic synthesis and analysis respectively (including the FFT).

\begin{figure}
\centering
\includegraphics[width=0.65\textwidth]{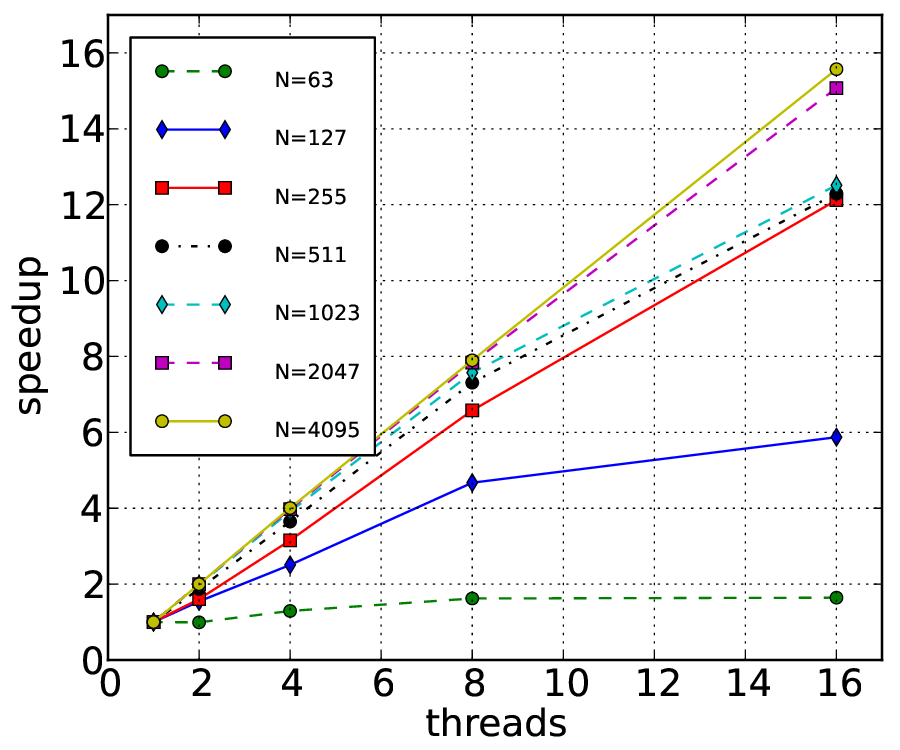}
\caption{Speedup obtained with multiple threads using OpenMP (\texttt{gcc} 4.6.3) on a 16 core intel Xeon E5-2680 (\emph{Sandy Bridge} architecture with AVX instruction set running at 2.7 GHz).}
\label{fig:avx_omp}
\end{figure}

Figure \ref{fig:avx_omp} shows the speedup $T(1)/T(p)$, where $T(p)$ is the largest of $T_s(p)$ and $T_a(p)$, and $T(1)$ is the time of the fastest single threaded tranform.
It shows that there is no point in doing a parallel transform with $N$ below 128. The speedup is good for $N=255$ or above, and excellent up to 8 threads for $N \ge 511$ or up to 16 threads for very large transform ($N \ge 2047$).

\subsection{Performance comparisons}

Table \ref{tab:speed} reports the timing measurements of two SHT libraries, compared to the optimized Gauss-Legendre implementation found in the \texttt{SHTns} library (this work).
We compare with the Gauss-Legendre implementation of \texttt{libpsht} \citep{libpsht}, a parallel spherical harmonic transform library targeting very large $N$, and with \texttt{SpharmonicKit} 2.7 (DH) which implements one of the Driscoll-Healy fast algorithms \citep{healy03}.
All the timings are for a complete SHT, which includes the Fast Fourier Transform.
Note that the Gauss-Legendre algorithm is by far (a factor of order 2) the fastest algorithm of the \texttt{libpsht} library.
Note also that \texttt{SpharmonicKit} is limited to $N+1$ being a power of two, requires $2(N+1)$ latitudinal colocation points, and crashed for $N=2047$.
The software library implementing the fast Legendre transform described by \citet{Mohlenkamp99}, \texttt{libftsh}, has also been tested, and found to be of comparable performance to that of \texttt{SpharmonicKit}, although the comparison is not straightforward because \texttt{libftsh} did not include the Fourier Transform.
Again, that fast library could not operate at $N=2047$ because of memory limitations.
Note finally that these measurements were performed on a machine that did not support the new AVX instruction set.

\begin{table*}
  \centering
	\begin{tabular}{c|ccccccc}		% resultats sur calcul3 (s2kit testé également : moins rapide que Healy).
           $N$                      & 63      & 127 & 255 & 511 & 1023 & 2047 & 4095 \\ \hline
           \texttt{libpsht} (1 thread) & 1.05 ms & 4.7 ms  & 27 ms  & 162 ms & 850 ms & 4.4 s & 30.5 s \\
           DH (fast)                & 1.1 ms  & 5.5 ms  & 21 ms  & 110 ms & 600 ms & NA    & NA \\
           \texttt{SHTns} (1 thread)  & 0.09 ms & 0.60 ms & 4.2 ms & 28 ms  & 216 ms & 1.6 s & 11.8 s
	\end{tabular}
	\caption{Comparison of execution time for different SHT implementations. The numbers correspond to the average execution time for forward and backward scalar transform (including the FFT) on an Intel Xeon X5650 (2.67GHz) with 12 cores. The programs were compiled with \texttt{gcc 4.4.5} and \texttt{-O3 -march=native -ffast-math} compilation options.}
	\label{tab:speed}
\end{table*}

\begin{figure}
\centering
\includegraphics[width=0.6\textwidth]{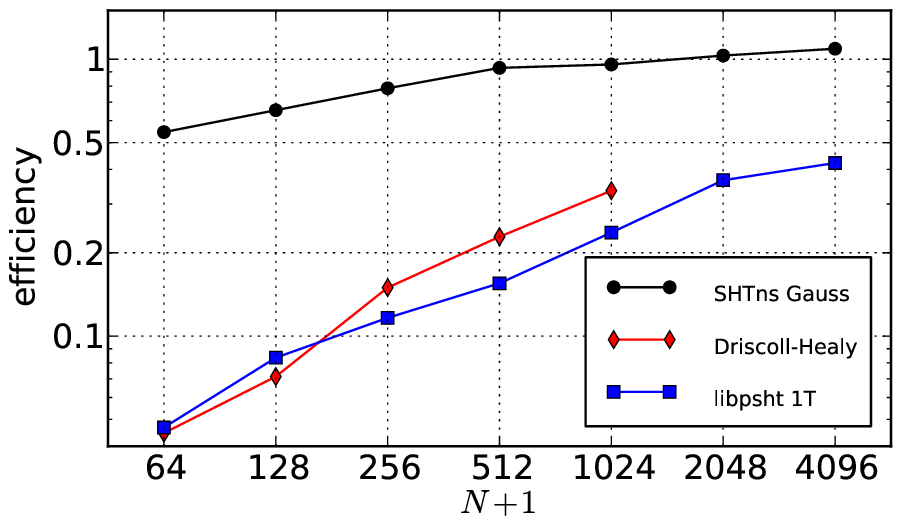}
\caption{Efficiency $(N+1)^3/(2Tf)$ of the implementations from Table \ref{tab:speed}, where $T$ is the execution time and $f$ the frequency of the Xeon X5650 CPU (2.67GHz) with 12 cores.}
\label{fig:speed}
\end{figure}

In order to ease the comparison, we define the efficiency of the SHT by $(N+1)^3/(2Tf)$, where $T$ is the execution time (reported in Table \ref{tab:speed}) and $f$ the frequency of the CPU.
Note that $(N+1)^3/2$ reflects the number of computation elements of a Gauss-Legendre algorithm (the number of modes $(N+1)(N+2)/2$ times the number of latitudinal points $N+1$). An efficiency that does not depend on $N$ corresponds to an algorithm with an execution time proportional to $N^3$.

The efficiency of the tested algorithms are displayed in Figure \ref{fig:speed}.
Not surprisingly, the Driscoll-Healy implementation has the largest slope, which means that its efficiency grows fastest with $N$, as expected for a fast algorithm.
It also performs slightly better than \texttt{libpsht} for $N \geq 511$.
However, even for $N=1023$ (the largest size that it can compute), it is still 2.8 times slower than the Gauss-Legendre algorithm implemented in \texttt{SHTns}.
It is remarkable that \texttt{SHTns} achieves an efficiency very close to 1, meaning that almost one element per clock cycle is computed for $N=\geq 511$.
Overall, \texttt{SHTns} is between two and ten times faster than the best alternative.
% We expect this gap to grow with the 4-vectors of the new AVX machines.

\subsection{Accuracy}

One cannot write about an SHT implementation without addressing its accuracy.
The Gauss-Legendre quadrature ensures very good accuracy, at least on par with other high quality implementations.

The recurrence relation we use (see \S\ref{sec:fly}) is numerically stable, but for $N \gtrsim 1500$, the value $P_m^m(x)$ can become so small that it cannot be represented by a double precision number anymore.
To avoid this underflow problem, the code dynamically rescales the values of $P_n^m(x)$ during the recursion, when they reach a given threshold.
The number of rescalings is stored in an integer, which acts as an enhanced exponent.
Our implementation of the rescaling does not impact performance negatively, as it is compensated by dynamic polar optimization: these very small values are treated as zero in the transform (eq. \ref{eq:synth_direct} and \ref{eq:gauss}), but not in the recurrence.
This technique ensures good accuracy up to $N=8191$ at least, but partial transforms have been performed successfully up to $N=43600$.

\begin{figure}
\centering
\includegraphics[width=0.5\textwidth]{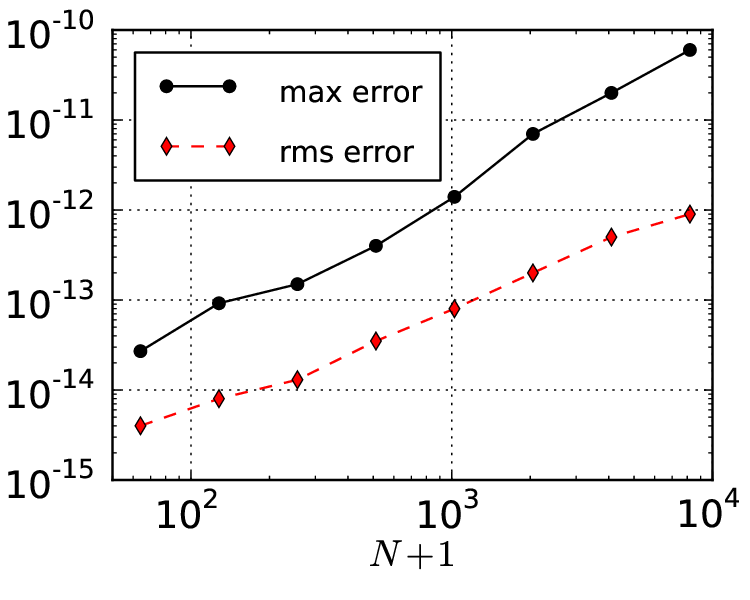}
\caption{Accuracy of the on-the-fly Gauss-Legendre algorithm with the default polar optimization.}
\label{fig:accuracy}
\end{figure}

To quantify the error we start with random spherical harmonic coefficients $Q_n^m$ with each real part and imaginary part between $-1$ and $+1$.
After a backward and forward transform (with orthonormal spherical harmonics), we compare the resulting coefficients $R_n^m$ with the originals $Q_n^m$.
We use two different error measurements: the maximum error is defined as 
$$ \epsilon_{max} = \max_{n,m} \left|R_n^m - Q_n^m\right| $$
while the root mean square (rms) error is defined as
$$ \epsilon_{rms} = \sqrt{\frac{2}{(N+1)(N+2)} \: \sum_{n,m} \left| R_n^m - Q_n^m \right|^2 } $$
The error measurements for our on-the-fly Gauss-Legendre implementation with the default polar optimization and for various truncation degrees $N$ are shown in Figure \ref{fig:accuracy}.
The errors steadily increase with $N$ and are comparable to other implementations.
For $N<2048$ we have $\epsilon_{max} < 10^{-11}$, which is negligible compared to other sources of errors in most numerical simulations.

\section{Conclusion and perspectives}

Despite the many fast spherical harmonic transform algorithms published, the few with a publicly available implementation are far from the performance of a carefully written Gauss-Legendre algorithm, as implemented in the \texttt{SHTns} library, even for quite large truncation ($N=1023$).
Explicitly vectorized on-the-fly algorithms seem to be able to unleash the computing power of nowadays and future computers, without suffering too much of memory bandwidth limitations, which is an asset for multi-threaded transforms.

The \texttt{SHTns} library has already been used in various demanding computations \citep[eg.][]{schaeffer2012, augier2013, figueroa2013}.
The versatile truncation, the various normalization conventions supported, as well as the scalar and vector transform routines available for C/C++, Fortran or Python, should suit most of the current and future needs in high performance computing involving partial differential equations in spherical geometry.

Thanks to the significant performance gain, as well as the much lower memory requirement of vectorized on-the-fly implementations, we should be able to run spectral geodynamo simulations at $N=1023$ in the next few years.
Such high resolution simulations will operate in a regime much closer to the dynamics of the Earth's core.

\section*{Acknowledgements}
The author thanks Alexandre Fournier and Daniel Lemire for their comments that helped to improve the paper.
Some computations have been carried out at the Service Commun de Calcul Intensif de l'Observatoire de Grenoble (SCCI) and other were run on the PRACE Research Infrastructure \emph{Curie} at the TGCC (grant PA1039).

\bibliography{sht}

\end{document}